\documentclass[11pt]{article}
\baselineskip
\usepackage{amssymb}
\usepackage{graphicx}
\usepackage{latexsym,cite}
\usepackage{epsfig}
\usepackage{psfrag}
\usepackage{amsmath, bbm, color}

\newcommand{\SU}{\mathrm{SU}}
\newcommand{\Z}{\mathbb{Z}}
\newcommand{\re}{{\rm{Re}}}
\newcommand{\dd}{{\rm{d}}}
\newcommand{\Tr}{{\rm Tr\,}}
\newcommand{\nconf}{n_{\mbox{\tiny{conf}}}}
\newcommand{\Se}{S^{\tiny\mbox{E}}}
\newcommand{\Sel}{S^{\tiny\mbox{E}}_{\tiny\mbox{L}}}
\newcommand{\Zlat}{Z_{\tiny\mbox{L}}}
\newcommand{\eq}{\begin{equation}}
\newcommand{\en}{\end{equation}}

\begin{document}

\begin{titlepage}
\vskip0.5cm 
\begin{flushright} 
DFTT 7/11\\  
HIP-2011-11/TH\\ 
\end{flushright} 
\vskip0.5cm 
\begin{center}
{\Large\bf
Thermodynamics of $\SU(N)$ Yang-Mills theories in $2+1$ dimensions I -- The confining phase
}
\end{center}
\vskip1.3cm
\centerline{Michele~Caselle$^{a}$, Luca~Castagnini$^{b}$, Alessandra~Feo$^{a}$, Ferdinando~Gliozzi$^{a}$ and Marco~Panero$^{c}$}
\vskip1.5cm
\centerline{\sl  $^a$ Dipartimento di Fisica Teorica dell'Universit\`a di Torino and INFN, Sezione di Torino,}    
 \centerline{\sl Via P.~Giuria 1, I-10125 Torino, Italy}
\vskip0.5cm
\centerline{\sl  $^b$ Institute for Theoretical Physics,  University of Regensburg,}    
 \centerline{\sl D-93040 Regensburg, Germany}
\vskip0.5cm
\centerline{\sl  $^c$ Department of Physics and Helsinki Institute of Physics, University of Helsinki,}    
 \centerline{\sl FIN-00014 Helsinki, Finland}
\vskip0.5cm
\begin{center}
{\sl  E-mail:} \hskip 5mm \texttt{caselle@to.infn.it, luca.castagnini@physik.uni-regensburg.de, feo@to.infn.it, gliozzi@to.infn.it, marco.panero@helsinki.fi}
\end{center}
\vskip1.0cm
\begin{abstract}
We compute the equation of state in the confining phase of $\SU(N)$ Yang-Mills theories with $N=2$, $3$, $4$, $5$ and $6$ colors in $2+1$ dimensions, via lattice simulations. At low enough temperatures, the results are accurately described by a gas of glueballs, including all known states below the two-particle threshold. Close to the deconfinement temperature, however, this prediction underestimates the numerical results, and the contribution from heavier glueballs has to be included. We show that the spectral density of the latter can be accurately described using a simple bosonic string model.
\end{abstract}
\vspace*{0.2cm}

\end{titlepage}

\section{Introduction and motivation} 
\label{introsect} 

Determining the phase diagram of strongly interacting matter is a major challenge in elementary particle physics, both theoretically and experimentally. 

From the theoretical point of view, the qualitative expectation that, when the temperature or the density is sufficiently high, usual hadronic matter gives way to a state of deconfined particles, is a straightforward consequence of asymptotic freedom, and has been around since the early days of QCD~\cite{first_deconfinement_prediction}. However, a derivation of the quantitative details of the QCD phase diagram is complicated by the fact that perturbative methods in thermal gauge theories are typically hindered by severe infrared divergences~\cite{Linde_problem}, and cannot be reliably applied close to the deconfinement point, where the physical coupling is not small. This leaves numerical computations based on the lattice regularization of QCD as the main tool for a first-principle study of the QCD phase structure as a function of the temperature $T$, and for vanishing or small values of the quark chemical potential $\mu$: recent results are summarized in refs.~\cite{lattice_QCD_EoS}.

On the other hand, on the experimental side, the creation of a deconfined plasma of quarks and gluons (QGP) in the laboratory has been the goal of a three-decade-long programme of heavy ion collision experiments, first at AGS and SPS, then at RHIC, and currently at LHC. In particular, the SPS, RHIC and LHC runs have provided convincing evidence for the creation of a state of deconfined matter, which achieves rapid thermalization, and can be characterized as a nearly ideal fluid~\cite{finiteTexperiments}.

It is important to point out that, in these experiments, the physical features of the deconfined plasma are studied \emph{indirectly}, namely, they are reconstructed from the properties (yields, momentum distributions et c.) of the hadrons produced after the expansion and freeze-out of the ``fireball''. The statistical analysis of these results is based on the assumption that, below the characteristic temperature range where deconfinement takes place (approximately between $150$ and $190$~MeV), thermal QCD can be modelled as a gas of massive, non-interacting hadronic resonances~\cite{Andronic:2009qf}. In fact, the very idea of a deconfinement temperature, as the limiting upper temperature at which the exponential growth of the density of states in the hadronic spectrum would lead to a divergence of the partition function, is even older than QCD~\cite{Hagedorn}.

Since hadrons are intrinsically non-perturbative objects, any first-principle test of the thermodynamic description of the confining QCD phase via the hadron resonance gas model necessarily requires lattice simulations. It should be noted that this is a computationally challenging task, because in the confining phase all equilibrium thermodynamic observables (such as pressure, energy density and entropy density) take values, which are much smaller than in the deconfined phase. However, the steady increase in computer power and major algorithmic improvements have now driven lattice QCD into an era of precision calculations, making it possible to reliably investigate the fine details of physical observables, even with limited computational resources.

Moreover, the most demanding technical aspects in lattice QCD computations involve the regularization of fermions, and thus can be easily bypassed, by restricting one's attention to the pure-glue sector---which captures most of the features of the full theory, at least at the qualitative or semi-quantitative level. The hadronic spectrum of pure Yang-Mills theories has been investigated extensively in highly accurate lattice computations~\cite{glueballs, Johnson:2000qz}, and several low-lying states are by now well-known: in particular, the lightest state in the $\SU(3)$ spectrum is a glueball with quantum numbers $J^{PC}=0^{++}$ and mass (in physical units) about $1.4$~GeV---significantly heavier than the lightest mesons in the physical QCD spectrum. The masses of glueballs with different quantum numbers are also known, and the most recent lattice calculations provide precise results for several excited states, too.\footnote{While the extraction of the latter involves a rapidly increasing computational complexity, the spectral density of glueball states at higher energies is expected to be approximately described by effective models~\cite{Isgur:1984bm}, based on the picture of glueballs as closed rings of glue.}

Looking at pure Yang-Mills theories also offers the further advantage of a cleaner symmetry pattern: as opposed to the full-QCD setup, global transformations associated with the center of the gauge group are an \emph{exact} symmetry of the Lagrangian, whose spontaneous breakdown can be studied by looking at the expectation value of the associated order parameter, namely the trace of the Polyakov loop. The latter provides an unambiguous definition of the deconfinement temperature $T_c$, which separates the center-symmetric, confining phase at low temperatures, from the deconfined phase at high temperatures.

The restriction to the Yang-Mills sector is also relevant in the 't~Hooft limit of QCD, i.e. in the double limit when the number of colors $N$ tends to infinity, and the coupling $g^2$ tends to zero, with the 't~Hooft coupling $\lambda=g^2N$ fixed~\cite{'tHooftlargeN}. In this limit, elementary combinatorics arguments show that quark-dynamics effects are subleading (more precisely: suppressed by powers of $N^{-1}$) with respect to contributions involving gluons only, and generic amplitudes for physical processes can be rearranged in double series, in powers of the 't~Hooft coupling $\lambda$, and of $N$---revealing a striking similarity to an analogous expansion in closed string theory: see, e.g., ref.~\cite{Aharony:1999ti} for a discussion. While these observations date back to more than thirty years ago, it is interesting to note that the large-$N$ limit also plays a technically important role in more modern analytical approaches to strongly coupled systems, based on the conjectured correspondence between gauge and string theories~\cite{Maldacena_conjecture}: according to this correspondence, the string-theoretical dual of a gauge theory simplifies to its classical gravity limit, when both the 't~Hooft coupling and the number of colors in the gauge theory are taken to be large. 

Finally, the emergence of a Hagedorn-like spectrum (i.e. an exponential growth in the number of hadronic states, as a function of their mass) has been studied in the large-$N$ limit of QCD, in both $D=2+1$ and $3+1$ spacetime dimensions, in a very recent work~\cite{Cohen:2011yx}.

With these motivations, in this paper we report our investigation of the equation of state in the confining phase of Yang-Mills theories in $2+1$ dimensions: we compare our results with a hadron resonance gas, using the glueball masses directly extracted from lattice simulations~\cite{Teper:1998te}, as well as a bosonic string model for the hadronic spectrum. This also allows one to achieve a better understanding of the many non-trivial features of effective string models---see, e.g., ref.~\cite{Teper:2009uf} and references therein.

Our computations can be compared with those reported in ref.~\cite{Meyer:2009tq}  for $\SU(3)$ in $D=3+1$: in fact, our work can be seen as an extension (in the lower-dimensional case) of the latter study, to theories with a different number of colors. This is partially motivated by recent works~\cite{finiteTlargeNlatticeresults}, revealing that the equation of state of the deconfined gluon plasma is characterized by a very mild dependence on the number of colors---up to a (trivial) proportionality to the number of gluon degrees of freedom---, showing that equilibrium thermodynamic observables in the $\SU(3)$ theory~\cite{Boyd:1996bx, Borsanyi:2011zm} are close to the large-$N$ limit, and lending support to computations based on holographic methods~\cite{holographicmodels,  Kiritsis_model_and_related_models} and/or on quasiparticle approaches~\cite{quasiparticlemodels}. By contrast, for temperatures $T<T_c$, confinement into color-singlet hadrons leads to the expectation that the number of physical states (and the equilibrium thermodynamic quantities) should scale as $\mathcal{O}(N^0)$, i.e. be independent of $N$, in the large-$N$ limit. On the other hand, our choice to look at the $D=2+1$ setup (rather than $D=3+1$) is motivated by an important technical aspect: while the deconfinement phase transition of $\SU(3)$ Yang-Mills theory in $3+1$ dimensions is of first order, in $2+1$ dimensions it is a second-order one. As a consequence, it is expected that the Hagedorn temperature $T_H$ should be the same as the deconfinement temperature $T_c$, thereby removing the $T_H/T_c$ parameter to be fitted from the data, and providing a more stringent test of the description of the glueball spectral density through a string model. Another lattice study of the equation of state in $2+1$ dimensions (but for the $\SU(3)$ gauge theory only) is reported in ref.~\cite{Bialas:2008rk}.

The structure of this paper is the following: in section~\ref{2+1_gauge_theories_section}, we briefly recall the continuum formulation and most interesting physical features of $\SU(N)$ Yang-Mills theories in $2+1$ spacetime dimensions, define their lattice regularization, and summarize some basic technical information about our determination of the thermodynamic quantities on the lattice. In section~\ref{results_section}, we present the numerical results of our simulations, and compare them with the equation of state predicted for a gas of non-interacting glueballs, using the glueball masses known from lattice computations. In section~\ref{glueball_models_section}, we define the effective description of the glueball spectrum of $\SU(N)$ Yang-Mills theories in $2+1$ dimensions in the large-$N$ limit through a bosonic string model, and derive the corresponding predictions for the equilibrium thermodynamic quantities considered in this work. Finally, in section~\ref{conclusions_section} we discuss our findings and their implications. The computation of the partition function for an ideal relativistic Bose gas is reviewed in the appendix~\ref{bosegas}, while appendix~\ref{closedstring} reports the derivation of the spectral density for a bosonic closed string model. Preliminary results of this study have been presented in ref.~\cite{Caselle:2010qd}.

\section{Non-Abelian gauge theories in $2+1$ dimensions in the continuum and on the lattice} 
\label{2+1_gauge_theories_section}

In this section, we first introduce the continuum formulation of $\SU(N)$ Yang-Mills theories in $2+1$ dimensions in subsection~\ref{2+1_Yang_Mills_subsection}, then we discuss their lattice regularization in subsection~\ref{SUN_lattice_regularization_subsection}, which also includes some technical details about our computation of thermodynamic quantities.

\subsection{Formulation in the continuum}
\label{2+1_Yang_Mills_subsection}

Contrary to the $D=1+1$ case, $\SU(N)$ gauge theories in $D=2+1$ spacetime dimensions exhibit non-trivial dynamics, and share many qualitative features with Yang-Mills theories in $D=3+1$. They are formally defined through the following Euclidean functional integral:
\eq
\label{continuum_formulation}
Z = \int\mathcal{D} A e^{-\Se}, \;\;\; \Se = \int {\dd}^3x \frac{1}{2 g_0^2 }\Tr F_{\alpha\beta}^2.
\en
In $D=2+1$ dimensions, the bare square gauge coupling $g_0^2$ has energy dimension $1$, so that bare perturbation theory calculations at a momentum scale $k$ are organized as series in powers of the dimensionless ratio $g_0^2/k$~\cite{3d_YM_renormalization_properties}. Like in $D=3+1$ dimensions, also in $D=2+1$ non-Abelian gauge theories are asymptotically free at high energy, and confining, with a finite mass gap and a discrete spectrum, at low energy. Their phase diagram as a function of the temperature consists of a confined phase (with color-singlet physical states, which can be classified according to the irreducible representations of the $\mathrm{O}(2)$ group and charge conjugation) at low temperatures, and a deconfined phase at high temperatures. 

The deconfinement transition occurs at a finite critical temperature $T_c$, where the global $\Z_N$ center symmetry gets spontaneously broken, and the order parameter in the thermodynamic limit is the trace of the vacuum average Polyakov loop. In $D=2+1$ dimensions, the deconfinement transition turns out to be a second-order one for $\SU(2)$ and $\SU(3)$, while it is a very weakly first-order one for $\SU(4)$, and a stronger first-order one for $\SU(N \ge 5)$~\cite{SUN_thermodynamics_in_2_plus_1_dimensions, Liddle:2008kk}. 

Equilibrium thermodynamic quantities for $\SU(N)$ Yang-Mills theories in $D=2+1$ dimensions can be easily obtained from elementary thermodynamic identities. Let $Z(T,V)$ denote the partition function for an isotropic system of two-dimensional ``volume'' $V$ at temperature $T$; the free energy density $f$:
\begin{equation}
\label{free_energy}
f = -\frac{T}{V} \ln Z(T,V) \,,
\end{equation}
is related, in the thermodynamic limit, to the pressure $p$ via:
\begin{equation}
\label{p_equals_minus_f}
p = - \lim_{V \to \infty} f \, .
\end{equation}
In turn, the trace of the energy-momentum tensor $\Delta=T^\mu_{\phantom{\mu}\mu}$ is related to the pressure by:
\begin{equation}
\label{delta_and_p}
\frac{\Delta}{T^3} = T \frac{d}{d T} \left(\frac{p}{T^3}\right) 
\,,
\end{equation}
so that the energy and entropy densities (denoted as $\epsilon$ and $s$, respectively) can be expressed as: 
\begin{equation}
\label{energy_density}
\epsilon =  \Delta + 2p
\end{equation}
and
\begin{equation}
\label{entropy_density}
s = \frac{\Delta + 3p}{T} \, .
\end{equation}

\subsection{Lattice regularization}
\label{SUN_lattice_regularization_subsection}

In this work, we studied non-perturbatively theories based on $\SU(N)$ gauge groups with $N=2$, $3$, $4$, $5$ and $6$ colors, by regularizing them on a finite, isotropic cubic lattice $\Lambda$. In the following, let $a$ denote the lattice spacing and $L_s^2 \times L_t= (N_s^2  \times N_t )a^3$ the lattice volume. The lattice formulation regularizes the functional integral in eq.~(\ref{continuum_formulation}), trading it for the finite-dimensional multiple integral:
\eq
\label{lattice_partition_function}
\Zlat = \int \prod_{x \in \Lambda} \prod_{\alpha=1}^3 {\dd}U_\alpha(x) e^{-\Sel},
\en
where ${\dd}U_\alpha(x)$ is the Haar measure for each $U_\alpha(x) \in \SU(N)$ link matrix, and $\Sel$ denotes the standard Wilson lattice gauge action:
\eq
\label{Wilson_lattice_gauge_action}
\Sel= \beta \sum_{x \in \Lambda} \sum_{1 \le \alpha < \beta \le 3} \left[1 - \frac{1}{N} \re\Tr U_{\alpha\beta}(x)\right], \;\;\; \mbox{with:} \;\;\; \beta=\frac{2N}{g_0^2 a},
\en
where:
\eq
\label{plaquette}
U_{\alpha\beta}(x) = U_\alpha(x) U_\beta(x+a\hat\alpha) U^\dagger_\alpha(x+a\hat\beta) U^\dagger_\beta(x).
\en
Expectation values of gauge-invariant physical observables $O$ are defined by:
\eq
\label{expectation_value}
\langle O \rangle = \frac{1}{\Zlat} \int \prod_{x \in \Lambda} \prod_{\alpha=1}^3 {\dd}U_\alpha(x) \; O \; e^{-\Sel}
\en
and can be estimated numerically, via Monte Carlo sampling over a finite set of $\{ U_\alpha(x) \}$ configurations; in the following, the number of configurations used in our computations is denoted by $\nconf$. 

\begin{table}
\begin{center}
\begin{tabular}{| c | c | c | c | c | c |}
\hline
$N$  & $N_s^2 \times N_t$ & $n_\beta$ & $\beta$-range 
& $\nconf$ at $T=0$ & $\nconf$ at finite $T$ \\
\hline \hline
$2$ & $48^3$          &  $81$ & $[7.97,10.97]$ & $1 \times 10^5 $ & ---  \\ 
    & $90^2 \times 6$ &       &                & --- & $1 \times 10^5 $  \\ 
    & $56^3$          &  $81$ & $[9.235,12.735]$ & $1 \times 10^5 $ & ---  \\ 
   & $105^2 \times 7$ &       &                & --- & $1 \times 10^5 $  \\ 
    & $64^3$          &  $90$ & $[9.5,14.5]$  & $1 \times 10^5 $ & ---  \\ 
   & $120^2 \times 8$ &       &                & --- & $1 \times 10^5 $  \\ \hline
$3$ & $64^2 \times 8$ &  $29$ & $[15.0,20.0]$ & $1 \times 10^5 $ & $5 \times 10^5$  \\ 
    &                 &     & $[23.0,24.4]$ & $1 \times 10^5 $ & $8 \times 10^5$  \\ 
    &                 &     & $[24.6,32.0]$ & $1 \times 10^5 $ & $6 \times 10^5$  \\ \hline
$4$ & $48^2 \times 6$ & $161$ & $[30.0,46.0]$ & $2 \times 10^4$ & $1.6 \times 10^5 $ \\ 
    & $56^2 \times 7$ & $188$ & $[34.5,53.2]$ & $2 \times 10^4$ & $1.6 \times 10^5 $ \\  
    & $64^2 \times 8$ & $200$ & $[39.0,58.9]$ & $2.5 \times 10^4$ & $2 \times 10^5 $ \\ \hline
$5$ & $48^2 \times 6$ &  $24$ & $[51.0,53.2]$ & $1 \times 10^5 $ & $1 \times 10^5 $ \\ 
    &                 &     & $[54.0,58.5]$ & $1 \times 10^5 $ & $5 \times 10^5 $ \\
    &                 &     & $[59.0,60.0]$ & $1 \times 10^5 $ & $3 \times 10^5 $ \\
    &                 &     & $[61.0,64.0]$ & $1 \times 10^5 $ & $1 \times 10^5 $ \\ \hline
$6$ & $48^2 \times 6$ &  $16$ & $[75.0,78.0]$ & $1 \times 10^5 $ & $1 \times 10^5 $ \\ 
    &                 &     & $[79.0,83.0]$ & $1 \times 10^5 $ & $4 \times 10^5 $ \\ 
    &                 &     & $[84.0,86.0]$ & $1 \times 10^5 $ & $2 \times 10^5 $ \\ 
    &                 &     & $[88.0,95.0]$ & $1 \times 10^5 $ & $1 \times 10^5 $ \\
\hline
\end{tabular}
\end{center}
\caption{Parameters of the main set of lattice simulations used in this work: 
$N$ denotes number of colors, $N_t$ and $N_s$ are, respectively, the lattice sizes along the time-like and space-like directions (in units of the lattice spacing). $n_\beta$ denotes the number of $\beta$-values (i.e. of temperatures) that were simulated, in each $\beta_{min} \le \beta \le \beta_{max}$ 
interval; the $T=0$ and finite-$T$ statistics at each $\beta$-value are shown in the last two columns. For $N>2$, all $T=0$ simulations were performed on lattices of size $(aN_s)^3$.}
\label{parameters}
\end{table}

The numerical results presented in this work are based on sets of configurations (see table~\ref{parameters} for details) produced via a Markovian process with local updates; our code implements a combination of local heat-bath~\cite{heatbath} and overrelaxation steps~\cite{overrelaxation} on $\SU(2)$ subgroups~\cite{Cabibbo:1982zn}. For part of our simulations, we also used the Chroma suite~\cite{Edwards:2004sx}. To convert the lattice results obtained from simulations into physical quantities, one has to set the physical scale, i.e. to determine the value of the spacing $a$ as a function of the bare gauge coupling. This determination is done non-perturbatively, via the lattice computation of a reference quantity relevant for low-energy scales (such as, for example, the asymptotic slope $\sigma$ of the confining potential $V(r)$ between a pair of static sources at zero temperature at large distances $r$, or the critical deconfinement temperature $T_c$). In this work, the determination of the scale is done using lattice results available in the literature~\cite{Liddle:2008kk}, and is expressed by the following formula:
\eq
\label{scale_setting}
\frac{T}{T_c} = \frac{\beta - 0.22 N^2 + 0.5}{N_t \cdot \left( 0.357 N^2 + 0.13 - 0.211/N^2 \right)} \;.
\en
Essentially, the accuracy limits on this formula are set by the precision in the determination of the $T_c/\sqrt{\sigma}$ ratio in the continuum limit and in the large-$N$ limit from ref.~\cite{Liddle:2008kk}. The latter reports $0.9026(23)$ for the $N \to \infty$ limit of this ratio, with a finite-$N$ correction term proportional to $N^{-2}$ with coefficient $0.880(43)$ (this fit is shown to describe well the data, all the way down to $N=2$). As a consequence, the uncertainty on our determination of the temperature scale can be estimated to be of the order of $1\%$, and has a negligible impact on our analysis (for the sake of clarity, we omit the corresponding horizontal errorbars from our plots).

More generally, it is worth noting that, given that all numerical simulations are done at finite values of the spacing $a$, on the lattice different observables can be affected by different discretization artifacts, thus the choice of a particular observable to set the scale introduces a systematic uncertainty. However, the quantitative effect of such uncertainty is small, $\mathcal{O}(a^2)$; for a comparison with alternative definitions of the scale, see, e.g., refs.~\cite{Bialas:2008rk, Caselle:2004er}.

The lattice simulation of Yang-Mills theories in thermodynamic equilibrium is straightforward: the temperature (in natural units $\hbar=c=k_B=1$) is defined by the inverse of the lattice size, $T=1/(a N_t)$, along a compactified direction, with (anti-)periodic boundary conditions for bosonic (fermionic) fields, while the sizes in the other directions are kept sufficiently large, $N_s \gg N_t$, to enforce a good approximation of the thermodynamic limit---see, e.g., refs.~\cite{Gliozzi_finite_volume} for a discussion. To obtain the temperature dependence of all equilibrium thermodynamic quantities (or, more precisely, of their difference with respect to the value at $T=0$---for which we run simulations on lattices of size $(aN_s)^3$) we varied the temperature by changing $a$ (which is a function of $\beta$) at fixed $N_t$ and $N_s=8N_t$ (except for the $\SU(2)$ gauge group: see table~\ref{parameters} for details), and repeated the computations at increasing values of $N_t$ to estimate discretization effects and perform a continuum extrapolation. As compared to the so-called ``fixed-scale approach''~\cite{Umeda:2008bd}, this method allows one to perform efficiently a fine and accurate temperature scan. The trace of the energy-momentum tensor can be obtained from:
\eq
\label{Delta_lattice}
\Delta  = \frac{3}{a^3} \frac{\partial \beta}{\partial \ln a} \left( \langle U_\Box \rangle_{T} - \langle U_\Box \rangle_{0} \right),
\en
where $\langle U_\Box \rangle_{T}$ denotes the expectation value of the average plaquette at the temperature $T$, while the pressure can be determined using the ``integral method''~\cite{Engels:1990vr}:
\eq
\label{p_lattice}
p  = \frac{3}{a^3} \int_{\beta_0}^\beta \dd \beta^\prime \left( \langle U_\Box \rangle_{T} - \langle U_\Box \rangle_{0} \right),
\en
where $\beta_0$ is a value of the Wilson parameter corresponding to a temperature sufficiently deep in the confined phase. In the present work, the numerical evaluation of the integral in eq.~(\ref{p_lattice}) was done by the trapezoid rule, except for the $\SU(4)$ gauge theory, for which we used the method described by eq.~(A.4) in ref.~\cite{Caselle:2007yc}, which is characterized by systematic errors $\mathcal{O}(n_\beta^{-4})$. Thus, the uncertainty on the pressure depends on the statistical precision of the plaquette differences, and on the systematic uncertainty related to the choice of the lower integration extremum $\beta_0$. Since the plaquette differences at different values of $\beta$ are obtained from independent simulations, the statistical errors on $p$ are obtained using standard error propagation. As for the systematic uncertainty related to the choice of the lower integration extremum, we checked that, by virtue of the exponential suppression of the plaquette differences in the confined phase, pushing $\beta_0$ to even lower values than those we used, would not have any significant impact on our results for the pressure. The energy and entropy densities are then obtained from eq.~(\ref{energy_density}) and from eq.~(\ref{entropy_density}), respectively.

As a technical remark, note that eq.~(\ref{Delta_lattice}) and eq.~(\ref{p_lattice}) show that the determination of thermodynamic quantities from very fine lattices can be computationally rather demanding, given that they are extracted from differences of plaquette expectation values at zero and finite temperature, and such differences scale like $a^3$ (or $a^4$ in the $D=3+1$ case). However, our numerical results reveal a mild cutoff dependence (at least in the range of $N_t$ values that we simulated), allowing one to get a reliable extrapolation to the continuum limit. In particular, leading-order discretization terms affecting the Wilson lattice gauge action eq.~(\ref{Wilson_lattice_gauge_action}) are quadratic in $a$, so that continuum results can be obtained by extrapolation of fits in $1/N_t^2$ to the $N_t \to \infty $ limit. These results can then be compared with the theoretical predictions of effective models for the equation of state of Yang-Mills theories in the confined phase, as discussed below.

\section{Numerical results}
\label{results_section}

\begin{figure*}
\centerline{\includegraphics[width=.8\textwidth]{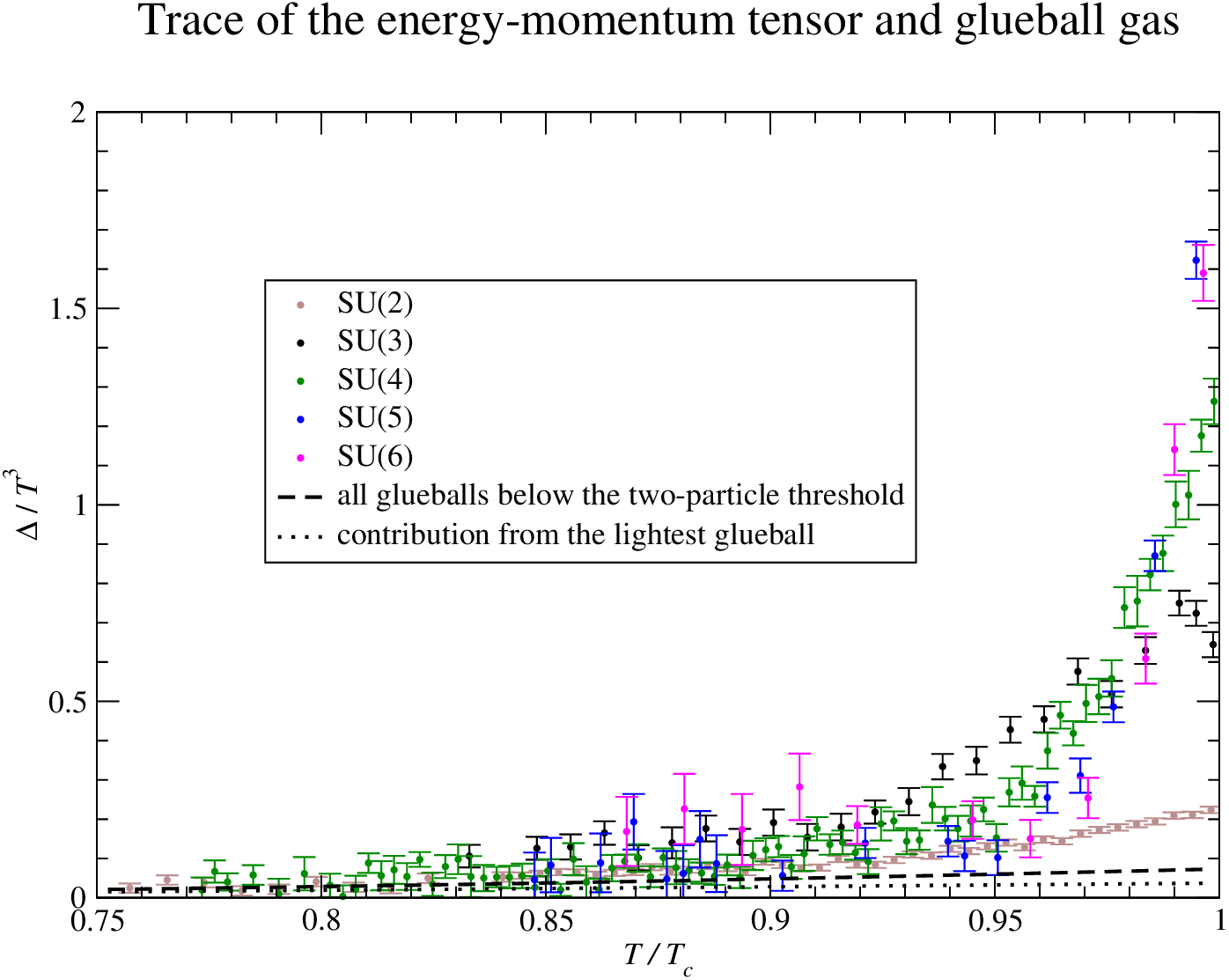}}
\caption{Trace of the energy-momentum tensor (in units of $T^3$) as a function of $T/T_c$, for the $\SU(N)$ gauge groups studied in this work. The results displayed are obtained from simulations on lattices with $N_t=6$ sites in the Euclidean time direction, except for the $\SU(3)$ Yang-Mills theory and for the points at the nine lowest temperatures of the $\SU(2)$ theory, obtained from simulations on lattices with $N_t=8$. The dashed curve is the theoretical prediction for $\Delta/T^3$, assuming that the system can be described as a gas of non-interacting glueballs, and the dotted curve represents the contribution from the lightest state in the spectrum.
}
\label{fig:Delta_lattice_glueballs}                              
\end{figure*}

Fig.~\ref{fig:Delta_lattice_glueballs} shows our numerical results for the dimensionless ratio of the trace of the energy-momentum tensor over the cube of the temperature, $\Delta/T^3$, as a function of $T/T_c$, for the various $\SU(N)$ gauge groups. In the confining phase, the thermodynamics is that of an ensemble of color-singlet states, and, since the number of the latter does not depend on $N$ (with the exception of $N=2$, for which, due to the (pseudo-)real nature of all irreducible representations of $\SU(2)$, there exist no states with charge conjugation quantum number $\mathcal{C}=-1$), it is reasonable to expect that the equilibrium observables should not depend strongly on the rank of the gauge group. This is indeed observed in fig.~\ref{fig:Delta_lattice_glueballs}, showing the approximate collapse of data from different groups onto a universal curve, up to temperatures around $0.95 T_c$, or even larger. In the same figure, we also show the comparison with the curve (dashed line) describing the trace of the energy-momentum tensor for a relativistic gas of massive, non-interacting glueballs, using the glueball masses extracted from lattice computations in ref.~\cite{Teper:1998te}, and restricting to states below the two-particle threshold; the leading contribution is given by the lightest glueball (dotted line). As it will be discussed in section~\ref{glueball_models_section}, both these curves severely underestimate the lattice results at temperatures larger than approximately $0.9 T_c$.

The plot also shows that, very close to the deconfinement transition, the data corresponding to different gauge groups start arranging themselves according to the multiplicity given by the number of gluon degrees of freedom in the deconfined phase, $\mathcal{O}(N^2)$. The fact that this already occurs for temperatures below (albeit close to) $T_c$ is likely due to residual finite-volume artifacts of the lattice simulations, which become particularly severe for second-order (or very weak first-order) phase transitions such as those of $\SU(2)$, $\SU(3)$ and $\SU(4)$. 

\begin{figure*}
\centerline{\includegraphics[width=.31\textwidth]{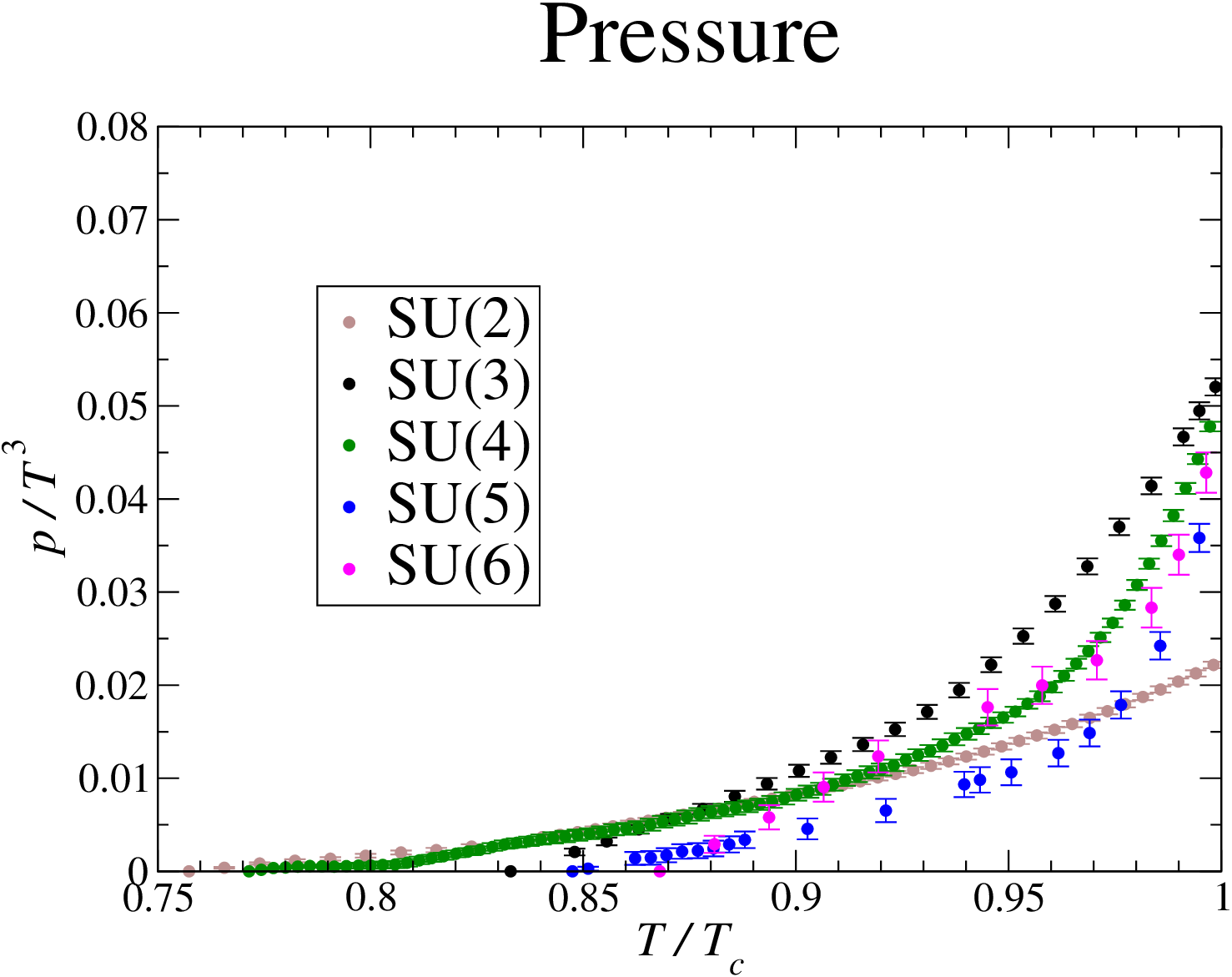} \hfill\includegraphics[width=.3\textwidth]{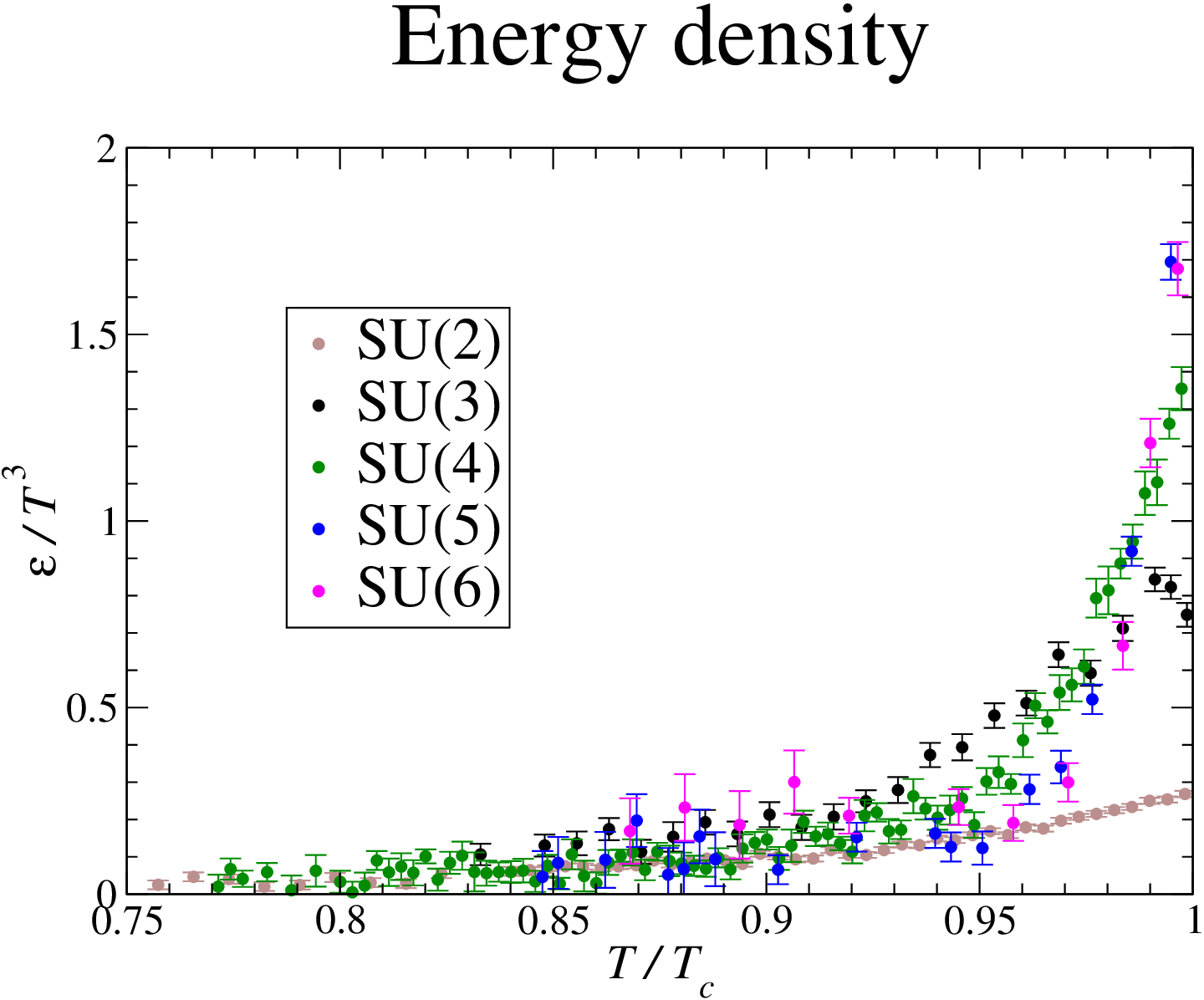} \hfill \includegraphics[width=.3\textwidth]{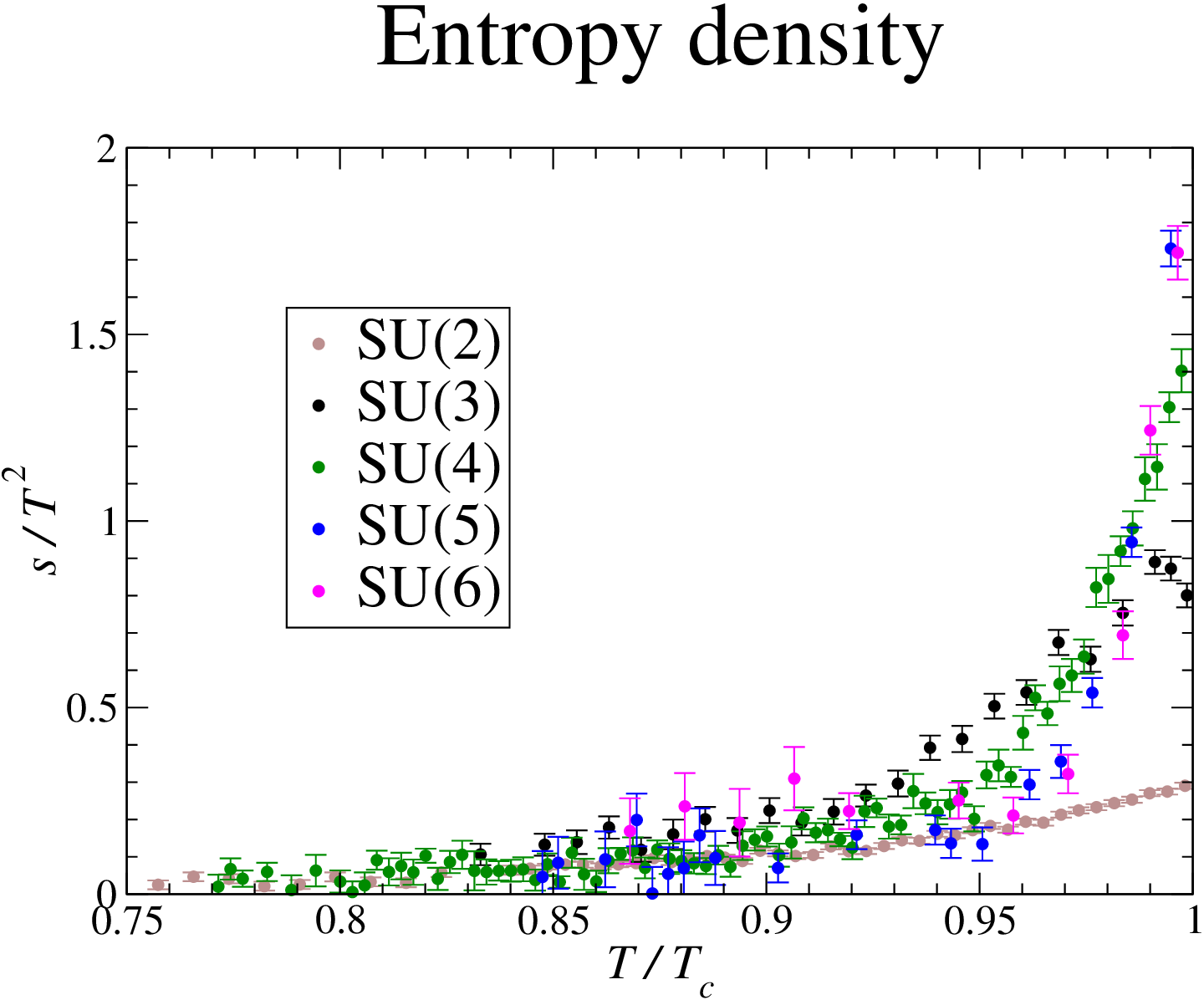}}                                                                     
\caption{The pressure (left panel) and the energy density (central panel), in units of $T^3$, and the entropy density (right panel) in units of $T^2$, as a function of $T/T_c$, for the theories studied in this work.}  
\label{fig:pressure_energy_entropy}                                                             \end{figure*}

From the data for the trace of the energy-momentum tensor, it is then straightforward to obtain the other bulk thermodynamic quantities $p$, $\epsilon$ and $s$ by numerical integration: the results are shown in fig.~\ref{fig:pressure_energy_entropy}, where the left, central, and right panels respectively show the pressure, the energy density and the entropy density (in units of the appropriate power of the temperature).

\begin{figure*}
\centerline{\includegraphics[width=.49\textwidth]{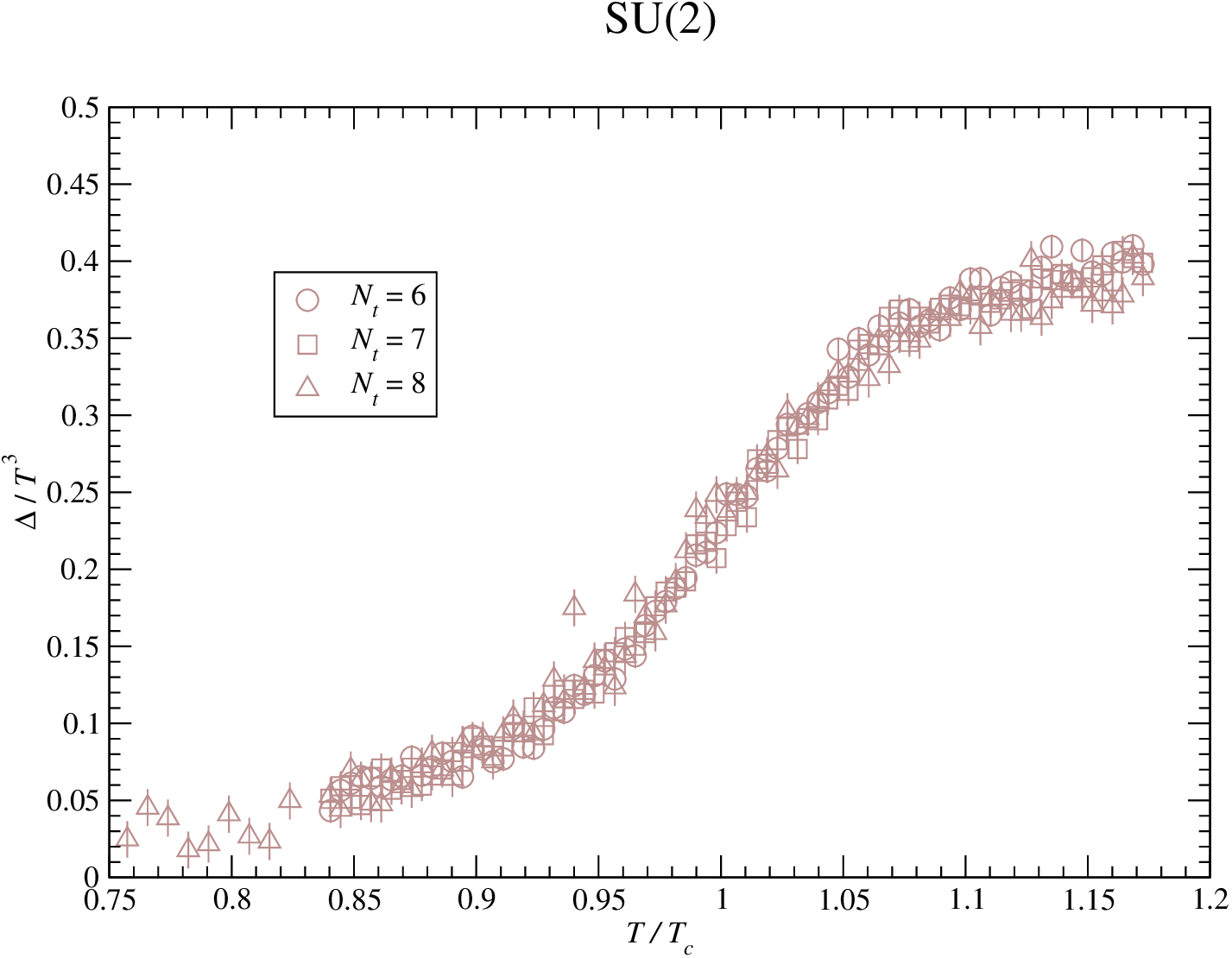} \hfill \includegraphics[width=.48\textwidth]{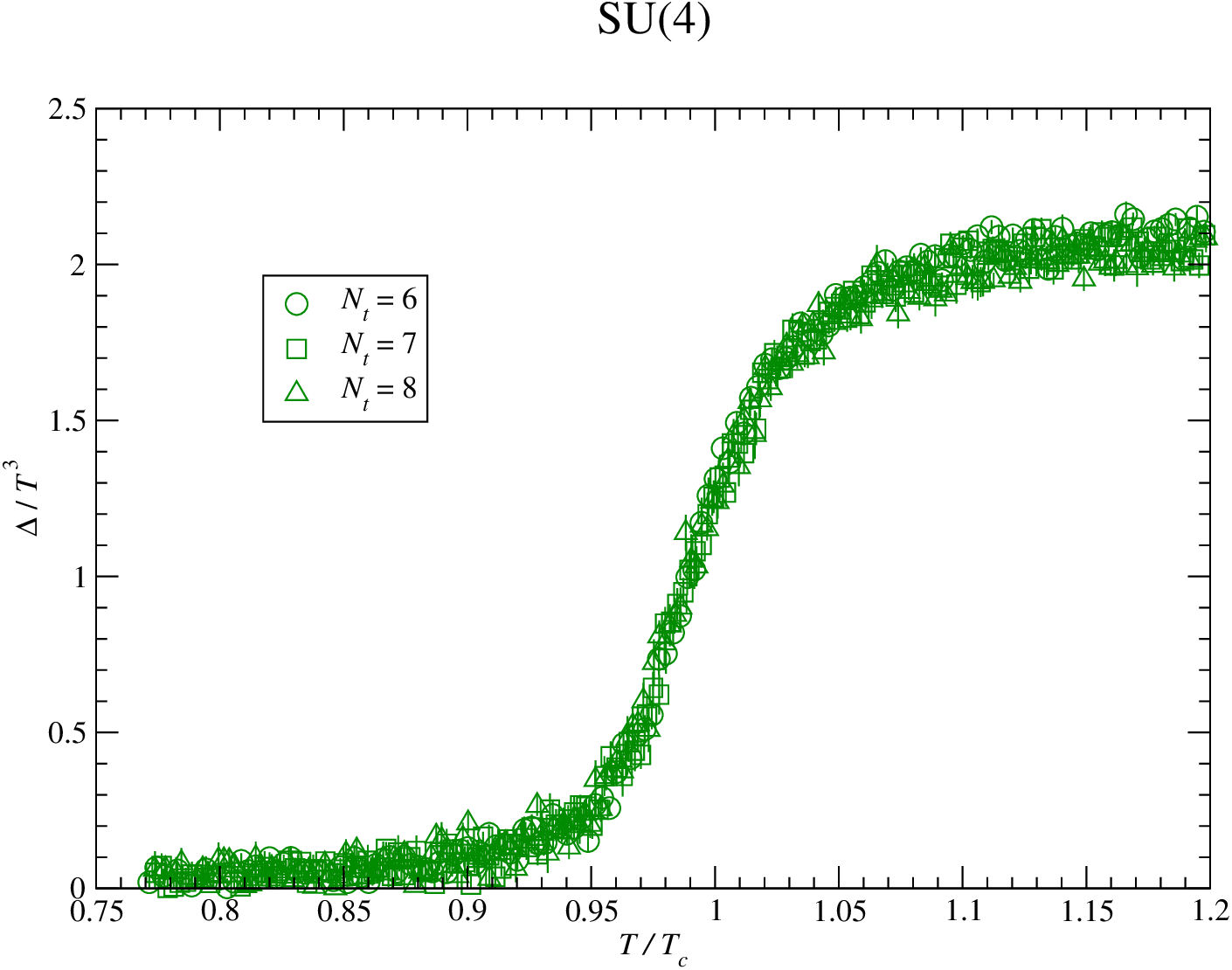}}
\vspace{1cm}
\caption{Cutoff dependence of our results for the trace of the energy-momentum tensor (in units of $T^3$). These panels show the results obtained at the same temperatures, from simulations at three different values of the spacing $a$, corresponding to lattices with $N_t=6$ (circles), $7$ (squares) and $8$ (triangles), for the $\SU(2)$ (left panel) and $\SU(4)$ gauge groups: the discretization effects appear to be comparable with or smaller than the statistical precision of our data.}
\label{fig:Nt_dependence}
\end{figure*}

Let us now discuss the continuum extrapolation of our lattice results. Since our numerical data are obtained from simulations of the pure-glue sector with the Wilson action eq.~(\ref{Wilson_lattice_gauge_action}), leading-order discretization effects are proportional to $a^{-2}$, i.e. to $N_t^{-2}$. To estimate the quantitative impact of deviations with respect to the continuum limit, we repeated our simulations of the $\SU(2)$ and $\SU(4)$ gauge theories on lattices with $N_t=6$, $7$ and $8$, keeping the temperature and physical volume fixed. Going from $N_t=6$ to $N_t=8$ is expected to reduce the dominating cutoff effects by a factor around one half, while keeping the computational costs at a constant signal-to-noise ratio limited. In fig.~\ref{fig:Nt_dependence} we compare the values for $\Delta/T^3$ for these two groups, as obtained from the three different sets of simulations: in the confined phase, discretization effects are very small, and compatible with the statistical errorbars. For this reason, one can safely assume that the systematic discretization effects affecting our $N_t=6$ results (as well as finite-volume effects) are negligible with respect to the statistical errors and systematic uncertainties related to the scale determination.

\section{Comparing with a glueball gas}
\label{glueball_models_section}

Since the models that we are studying are pure gauge theories, the only physical states in the confined phase are massive glueballs. Hence, as a first approximation, it is reasonable to expect that the behavior of the thermodynamic observables shown in fig.~\ref{fig:Delta_lattice_glueballs} and in fig.~\ref{fig:pressure_energy_entropy} could be described in terms of a free relativistic gas of these glueball states. It is far from obvious that these glueballs should behave as free particles (at least for small values of $N$), and testing this assumption is the first goal of our analysis. As we shall see, this also requires an \emph{Ansatz} for the glueball spectrum at high energies (in the vicinity of a possible Hagedorn-like transition), which, in turn, will allow us to discuss some subtle features of the phenomenological models used to describe this spectrum. Testing these string-inspired models is the second goal of our analysis.

We performed the comparison of our data to the ideal glueball gas predictions 
in three steps: 
\begin{enumerate}
\item firstly, we assumed the gas to be dominated by the lightest glueball only;
\item then, we included all the glueballs below the two-particle threshold, using the very precise numerical estimates available in the lattice literature;
\item finally, we compared our data with the whole glueball spectrum, assuming a spectral density \emph{Ansatz} inspired by the effective bosonic string model. 
\end{enumerate}
A similar approach has been followed in ref.~\cite{Meyer:2009tq} for the $\SU(3)$ Yang-Mills theory in $D=3+1$ dimensions. 

In $D=2+1$ spacetime dimensions, the pressure associated with a free, relativistic particle species of mass $m$ is 
\eq
\label{press}
p = \frac{m T^2}{2 \pi} \sum_{k=1}^{\infty} \frac{1}{k^2} \exp\left(-k \frac{m}{T}\right) \left(1 + \frac{T}{k m}
\right),
\en 
from which the other equilibrium thermodynamic quantities can be derived. In particular, the trace of the energy-momentum tensor can be written as 
\eq
\frac{\epsilon - 2 p}{T^3} = \frac{m^2}{2 \pi T^2} \sum_{k=1}^{\infty} 
\frac{1}{k} \exp\left(-k \frac{m}{T}\right)
=-\frac{m^2}{2\pi T^2}\ln\left(1-e^{-\frac{m}{T}}\right)
\label{new}
\en
(see also eq.~(\ref{identity}) in the appendix~\ref{bosegas}).

For the first two steps in the comparison of our lattice data to the glueball spectrum, we used the numerical values of the glueball masses and the parametrizations of the appropriate scaling functions, which are reported in ref.~\cite{Teper:1998te}. 

The curves in fig.~\ref{fig:Delta_lattice_glueballs} show the expected behavior of $\Delta/T^3$, for a gas of non-interacting glueballs: in particular, the dashed line is obtained summing the contributions from all glueball species (below the elastic scattering threshold) which are known from lattice spectroscopy calculations, while the dotted curve represents the leading contribution given by the lightest glueball. As already mentioned above, it is easy to see that both these curves fail to reproduce the data for $T/T_c$ larger than (approximately) $0.9$. This has also been observed in recent, high-precision lattice computations of the equation of state for $\SU(3)$ Yang-Mills theory in $D=3+1$ dimensions~\cite{Meyer:2009tq, Borsanyi:2011zm}.

Another feature, which is immediately manifest from the data, is the large separation between the bands of data corresponding to the $\SU(2)$ and the $\SU(N\geq3)$ gauge groups. Since the value of the lowest glueballs (in units of $T_c$) is almost the same for $\SU(2)$ and for the other $\SU(N \geq 3)$ gauge groups, this gap must be the consequence of the fact that in the two cases the 
theories have different spectra, due to the aforementioned absence of $\mathcal{C}=-1$ states in the $\SU(2)$ Yang-Mills theory. This is an important difference with respect to the spectrum of the theories based on the other $\SU(N \geq 3)$ gauge groups, admitting both $\mathcal{C}=+1$ and $\mathcal{C}=-1$ states (not mutually degenerate). The fact that our $\SU(2)$ results for $\Delta/T^3$ start to strongly deviate from those of the other groups at $T/T_c \simeq 0.9$ indicates that in this region the thermodynamics is likely dominated by effects due to the density of glueball states, rather than by just the lightest state in the spectrum.

To describe the full glueball spectrum, various phenomenological models have been proposed in the literature: these include, in particular, bag-type models~\cite{Karl_Paton} and string-inspired models~\cite{Isgur:1984bm}. In the following we focus on the latter, and summarize their main features; besides the original work, the interested readers can find a discussion of its more recent generalizations in ref.~\cite{Johnson:2000qz}. 

In the original proposal by Isgur and Paton~\cite{Isgur:1984bm}, glueballs are modelled as ``rings of glue'', i.e. as closed tubes of chromoelectric flux,
which are described as closed bosonic string states. In particular, this implies that each glueball state corresponds to a given phonon configuration (i.e. to a given bosonic closed string state), and for each phonon combination there exists an infinite tower of radially excited states of increasing mass. This model can then be generalized, by including possible $k$-glueball states (for $N \ge 4$), which correspond to closed $k$-strings, metastable ``adjoint string glueballs'' (which become stable in the large-$N$ limit and may explain the
splitting between the sectors of opposite $\mathcal{C}$), as well as the finite thickness of the flux tube, which is usually modelled introducing an additional 
phenomenological parameter. This generalized version of the Isgur-Paton model turns out to be in remarkably good agreement with the low-lying spectrum of Yang-Mills theories, as calculated from first principles by means of lattice simulations~\cite{Johnson:2000qz}.

An interesting feature of this model is that, essentially, these extensions lead to copies of the original spectrum, which are shifted towards higher values of the
masses: thus, the thermodynamic contribution of the corresponding states is exponentially suppressed, except in a close neighborhood of a Hagedorn-like temperature. Furthermore, the correction to the spectrum due to the finite thickness of the flux tube becomes negligible for heavy glueballs, so that, as a first approximation, the glueball spectral density can be modelled in terms of the spectrum of a closed bosonic string (see the appendix~\ref{closedstring} for details):
\eq
\label{string}
\tilde \rho_{D}(m) = \frac{ (D-2)^{{D}-1} }{m} \left( \frac{\pi T_H}{3m} \right)^{D-1} e^{m/T_H} \, .
\en
For $\SU(N \geq 3)$, the model predicts a further twofold degeneracy, accounting for the two possible orientations of the flux tube.

\begin{figure*}
\centerline{\includegraphics[width=.8\textwidth]{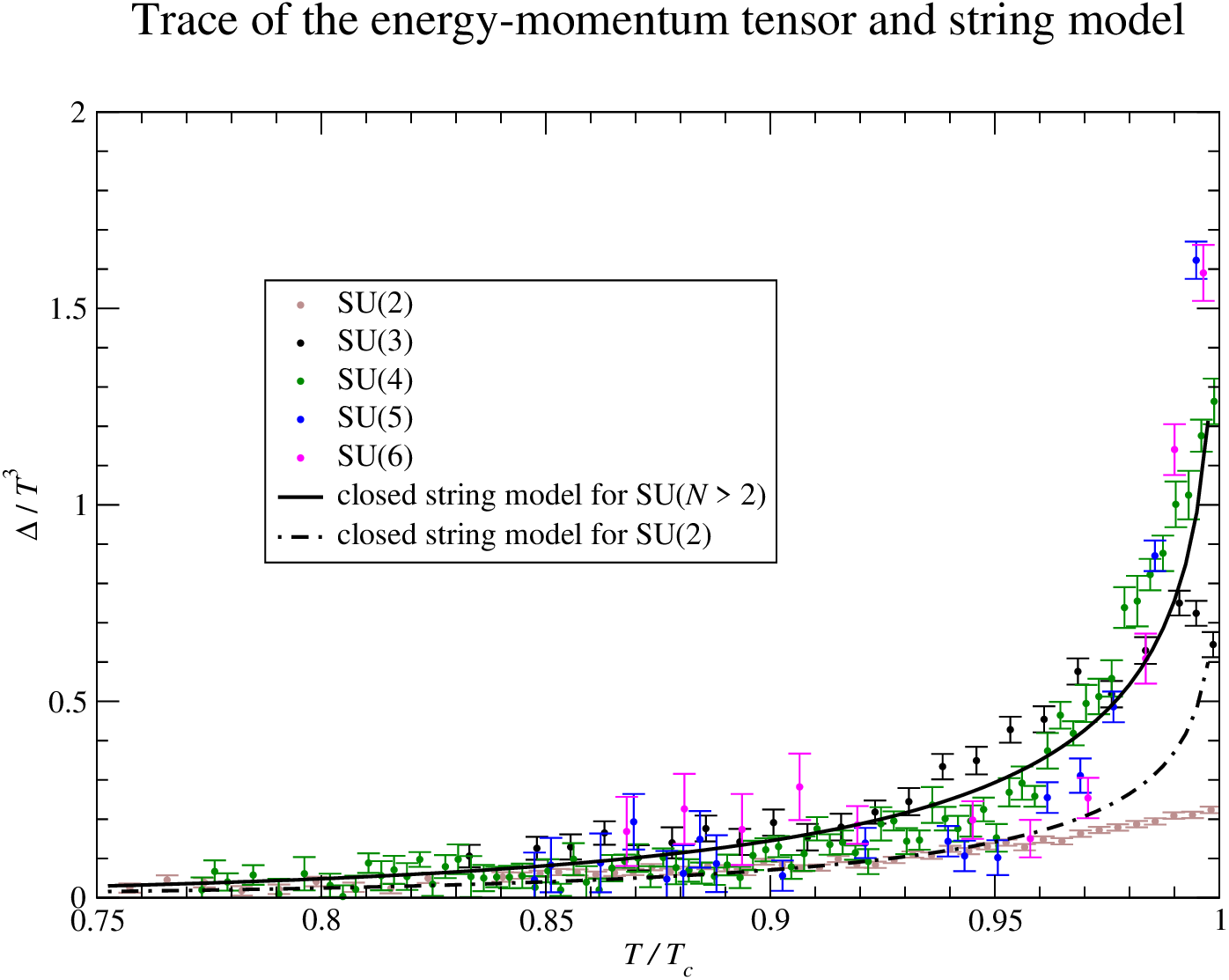}}
\caption{Same as in fig.~\protect\ref{fig:Delta_lattice_glueballs}, but comparing our simulation results to the theoretical prediction including the contribution obtained from a bosonic string model for the glueball spectral density, as discussed in the text. The solid curve is the prediction for the $\SU(N > 2)$ theories, while the dash-dotted curve accounts for the lack of $\mathcal{C}=-1$ states in the $\SU(2)$ gauge theory.
}
\label{fig:Delta_string_model}                              
\end{figure*}  
 
Using this expression in eq.~(\ref{new}), assuming $T_H=T_c$ and a Nambu-Goto string, for which $T_c^2=3\sigma/\pi$ in $D=2+1$ spacetime dimensions,\footnote{It is interesting to note that, in $D=2+1$ dimensions, the effective Nambu-Goto string model for confinement predicts a numerical value for the ratio of the deconfinement temperature over the square root of the zero-temperature string tension $T_c/\sqrt{\sigma}$ approximately equal to one, in good numerical agreement with recent, accurate lattice determinations~\cite{Liddle:2008kk, Bialas:2009pt}.} one obtains the prediction shown in fig.~\ref{fig:Delta_string_model}. In particular, the solid curve, which accounts for states of opposite charge-conjugation quantum number, is relevant for the trace anomaly in $\SU(N > 2)$ theories, while the dash-dotted curve, obtained including only $\mathcal{C}=+1$ states, is expected to provide a more adequate description for the $\SU(2)$ theory. As the figure shows, both curves are in remarkably good agreement with our data, for all temperatures up to the region where the results corresponding to the different gauge groups start splitting from each other. This agreement is a strong piece of numerical evidence supporting the Isgur-Paton model, and, more generally, bosonic string models as effective theories for the confining regime of non-Abelian gauge theories. These results also support the hypothesis that glueball interactions are weak, and that the confined phase thermodynamics can be accurately approximated in terms of a relativistic gas of free massive bosons.

\begin{figure*}
\centerline{\includegraphics[width=.8\textwidth]{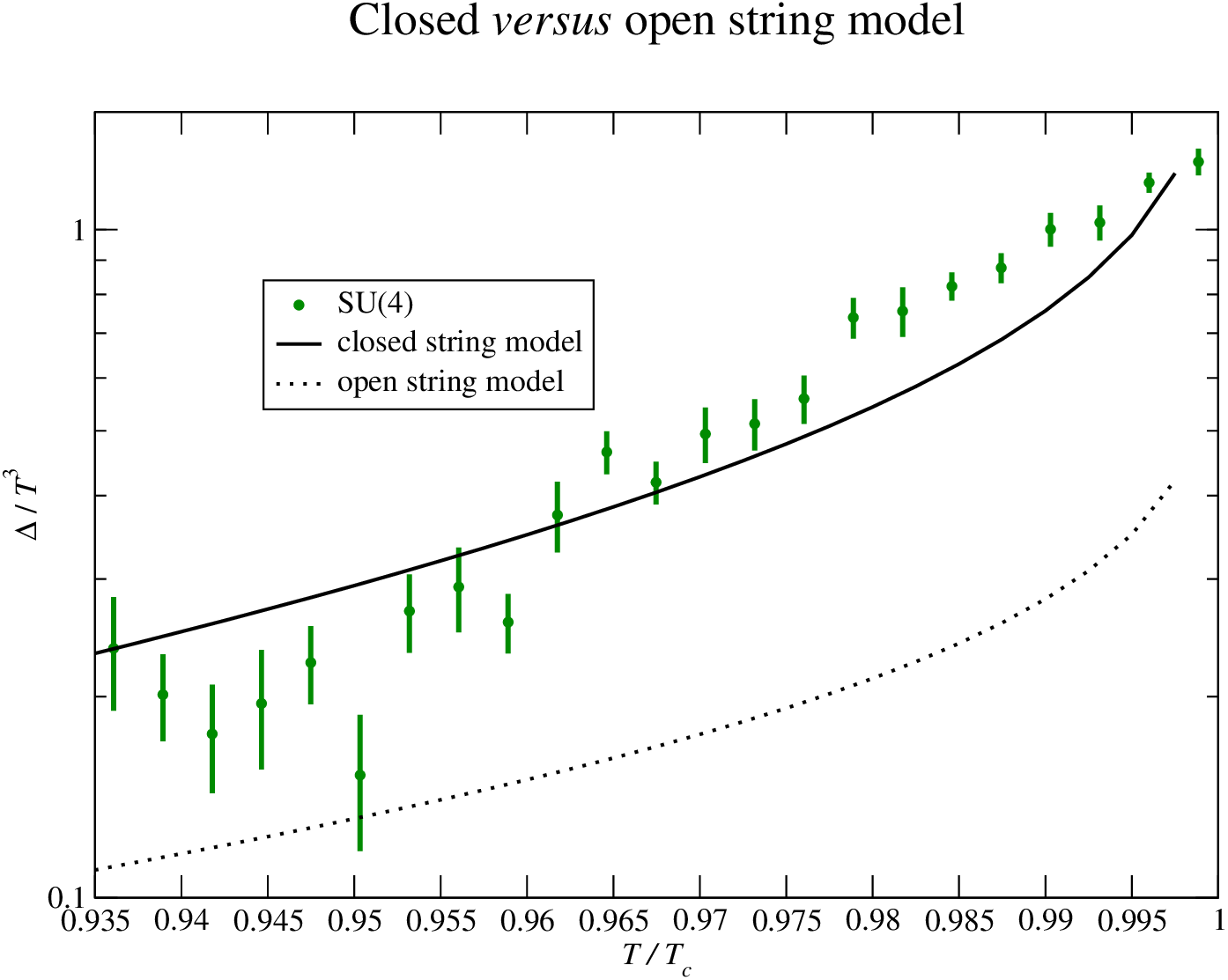}}
\caption{Dependence of the theoretical prediction for the equation of state on the string spectrum details: our simulation results for the $\SU(4)$ Yang-Mills theory are compared with the bosonic string prediction for the equation of state, assuming the glueballs to be modelled either as closed (solid curve) or open (dotted curve) strings.}
\label{fig:logarithmic_open_vs_closed_string_model}
\end{figure*}

In order to have a better feeling of the quality of the agreement between lattice data and the effective string prediction, it is also useful to compare our simulation results with the prediction that one would obtain, using the density of states of the \emph{open} (rather than \emph{closed}) string. This is displayed in fig.~\ref{fig:logarithmic_open_vs_closed_string_model}, where we compare the two curves to our data for $\SU(4)$ (the gauge group for which we performed the finest temperature scan) in the region where the contribution from the density of states of heavy glueballs dominates over the lightest ones: the precision of our simulations is sufficient to show that a model based \emph{only} on open strings is clearly incompatible with the lattice results. However, our work does not rule out the possibility of modelling the glueballs in terms of a combination of closed and open string states in the adjoint (or in a higher) representation, as suggested in ref.~\cite{Johnson:2000qz}. 

Finally, in the close vicinity of $T_c$, our results show that the contribution from heavier glueballs (or from interactions) becomes more and more important: this drives the change of behavior observed in the figures. It is interesting to note that, while the original Isgur-Paton model predicts exactly the same glueball spectrum for any number of colors $N$ (except for the missing $\mathcal{C}=-1$ states for $N=2$), the extension discussed in ref.~\cite{Johnson:2000qz} predicts a dependence on $N$, related to the larger number of $k$-glueball states which become available when $N$ is increased.

\section{Conclusions}
\label{conclusions_section}

In this work, we presented high-precision lattice results for the equation of state of $\SU(N)$ Yang-Mills theories in $2+1$ dimensions. We focused onto the confining phase, where the thermodynamics of these strongly coupled theories is expected to be described in terms of color-singlet hadronic states (glueballs). 

At low enough temperatures, the equilibrium thermodynamic properties are described well by a gas of non-interacting glueballs, with masses compatible with the results obtained from the accurate lattice determinations available in the literature.

Close to the deconfinement temperature, however, this very simple model fails to reproduce the lattice data, and the contribution due to heavier glueball states has to be taken into account. The latter can be evaluated using a simple bosonic string model, like the one originally proposed in ref.~\cite{Isgur:1984bm} for the $D=3+1$ case, or a refinement thereof~\cite{Johnson:2000qz}. The resulting equation of state, assuming that the Hagedorn temperature can be identified with the deconfinement temperature, and taking the effective string to be described by the Nambu-Goto model, is in very good agreement with the data from our lattice simulations. This gives further support to the validity of bosonic string models as effective theories for the confining phase in non-Abelian gauge theories.

Our findings can be compared with those obtained in a similar computation for the $\SU(3)$ Yang-Mills theory in $D=3+1$ dimensions~\cite{Meyer:2009tq}, which also reported excellent agreement with an equation of state obtained extending the sum over known glueball masses with an exponential Hagedorn-like spectrum. One difference with respect to the latter work, however, is that in the present work we did not fit the value of the Hagedorn temperature $T_H$: in the $D=2+1$ setup the deconfinement phase transition is a second-order one also for the $\SU(3)$ gauge theory---and a very weakly first-order one for $\SU(4)$---and for continuous phase transitions it is expected that $T_H$ should be equal to $T_c$.

\vskip1.0cm {\bf Acknowledgements.}\\
L.C. acknowledges partial support from Deutsche Forschungsgemeinschaft (Sonderforschungsbereich/Transregio 55) and the European Union grant 238353 (ITN STRONGnet). M.P. acknowledges financial support from the Academy of Finland, project 1134018. Numerical simulations were partially performed on the INFN Milano-Bicocca TURING cluster.

\appendix
\section{Ideal relativistic Bose gas in $d+1$ spacetime dimensions}
\label{bosegas}
\renewcommand{\theequation}{A.\arabic{equation}}
\setcounter{equation}{0}

The logarithm of the canonical partition function $Z(T,V)$ of an ideal relativistic 
Bose gas is:
\eq
\ln Z=-\frac {V \Omega_d}{(2\pi)^d}\,
\int_0^\infty \dd p~p^{d-1} \ln \left( 1-e^{-\sqrt{m^2+p^2}/T} \right)\,,
\en
where $m$ is the mass of the boson and $\Omega_d=2\pi^{d/2}/\Gamma(d/2)$ is the $d$-dimensional solid angle. Integration by parts yields:
\begin{eqnarray}
\ln Z &=& \frac{V \Omega_d}{Td(2\pi)^d}
\int_0^\infty \!\! \dd p \frac{p^{d+1}}{\sqrt{m^2+p^2}} \frac{1}{e^{\sqrt{m^2+p^2}/T} -1} \nonumber\\
 &=& \frac{V \Omega_d}{Td(2\pi)^d} \sum_{k=1}^\infty
\int_0^\infty \!\! \dd p \frac{p^{d+1}}{\sqrt{m^2+p^2}} e^{-k\sqrt{m^2+p^2}/T} \nonumber\\
&=& \frac{m^{d+1} V \Omega_d}{Td(2\pi)^d} \sum_{k=1}^\infty \int_0^\infty \dd u \; e^{-k\frac{m}{T}\cosh u} \; \sinh^{d+1} u  \nonumber \\
&=& \frac{2V}{T} \left( \frac{m^2}{2\pi} \right)^{\frac{d+1}{2}} \sum_{k=1}^\infty \left( \frac{T}{k m} \right)^{\frac{d+1}{2}} K_{\frac{d+1}{2}}\left(k\frac{m}{T}\right)\; ,
\end{eqnarray}
where we set
$\cosh u=\sqrt{1+\frac{p^2}{m^2}}$, and used the following integral representation:
\eq
K_\nu(z)=\frac{\sqrt{\pi}\left( \frac{z}{2} \right)^\nu}{\Gamma \left( \nu+ \frac{1}{2}\right) } \int_0^\infty \dd u \; e^{-z\cosh u} \; \sinh^{2\nu}u
\en
for the modified Bessel function of the second kind of index $\nu$.

In the thermodynamic limit the pressure is
\eq
p=\frac{T}{V}\ln Z=2 \left( \frac{m^2}{2\pi} \right)^{\frac{d+1}{2}}
\sum_{k=1}^\infty \left( \frac{T}{k m} \right)^{\frac{d+1}{2}}
K_{\frac{d+1}{2}} \left( k\frac{m}{T} \right)~.
\label{pressure_Bessel_sum}
\en
The other equilibrium thermodynamics observables can be obtained from the above expressions for the pressure. For instance, the entropy density $s$ is given by 
\eq
s=\frac{\partial p}{\partial T}\,.
\en
Similarly, the internal energy density $\epsilon$ reads: 
\eq
\epsilon=\frac{T^2}{V}\frac{\partial}{\partial T}\ln Z=-p+sT~.
\en
Combining eq.~(\ref{pressure_Bessel_sum}) with the expression for the  trace of the energy-momentum tensor in $d$ spatial dimensions, $\Delta_d=\epsilon-d \cdot p$, and using the recurrence relations of modified Bessel functions, one finds that the ideal Bose gas enjoys a remarkable identity:
\eq
\Delta_d = 2\left(\frac{m^2}{2\pi} \right)^{\frac{d+1}2}\sum_{k=1}^\infty \left( \frac{T}{k m} \right)^{\frac{d-1}{2}} K_{\frac{d-1}{2}} \left( k \frac{m}{T} \right)\,,
\en
namely, the trace of the energy-momentum tensor for the Bose gas in $d$ spatial dimensions is proportional to the pressure $p_{d-2}$ of a Bose gas in $d-2$ spatial dimensions:
\eq
\Delta_d = \frac{m^2}{2\pi}\,p_{d-2}~.
\label{identity}
\en
Finally, note that using the asymptotic expansion
\eq
K_\nu\simeq\sqrt{\frac{\pi}{2z}}e^{-z}\left[ 1+\frac{4\nu^2-1}{8z}
+\mathcal{O}\left(\frac{1}{z^2}\right)\right]\,,
\en
valid for large $| z |$, one obtains:
\eq
p\simeq T\left(\frac{Tm}{2\pi}\right)^{\frac{d}{2}} \sum_{k=1}^\infty
\frac{1}{k^{\frac{d}{2}+1}} \exp\left(-k \frac{m}{T}\right) \left[1+\frac{d(d+2)}{8k}\frac Tm\right]\,,
\en
which, for $d=2$, reduces to eq.~(\ref{press}).

\section{Spectral density of closed bosonic strings}
\label{closedstring}
\renewcommand{\theequation}{B.\arabic{equation}}
\setcounter{equation}{0}

We assume that the glueballs can be modelled as ``rings of glue'', which are described, in the limit of large masses, by the Nambu-Goto model of closed bosonic strings. The mass spectrum in $D=2+1$ spacetime dimensions reads:
\eq
m^2 = 4\pi \sigma \left( n_L + n_R - \frac{1}{12} \right)\;\;\; \mbox{with:}\;\;n_L=n_R=n\,,
\en
where $\sigma$ is the string tension, the integers $n_L$ and $n_R$ describe the total contribution of the left- and right-moving phonons along the closed string, and the $-1/12$ term arises from the zero-point energy contribution. The degeneracy of these single-particle states is given by the number of partitions of $n_L$ and $n_R$ (see, for instance, ref.~\cite{Zwiebach_book}), hence the total degeneracy $\rho(n)$ for the physical states is
\eq
\rho(n)=\pi(n_L)\,\pi(n_R)=\pi(n)^2\,,
\en
where $\pi(n)$ denotes the number of partitions of $n$, and can be calculated using the generating function
\eq
\prod_{k=1}^\infty \frac{1}{1-q^k}=\sum_{n=0}^\infty \pi(n)q^n~.
\en
For $n$ large, one can resort to the Ramanujan asymptotic formula:
\eq
\pi(n)\simeq\frac{1}{4n\sqrt{3}} \exp \left( \pi \sqrt{\frac{2n}{3}} \right) \,,
\en
which gives:
\eq
\rho(n)\simeq \frac{1}{48n^2} \exp \left( 2 \pi\sqrt{\frac{2n}{3}} \right)\,.
\en
In $D$ spacetime dimensions, the Hagedorn temperature $T_H$~\cite{Hagedorn} is related to the string tension by:
\eq
T_H=\sqrt{\frac{3\sigma}{\pi(D-2)}}\,,
\en
thus, for large $m$,
\eq
\frac{m}{T_H}=2\pi\sqrt{\frac{2(D-2)n}{3}}\,,
\label{MN}
\en
so in $D=2+1$ one gets:
\eq
\rho(n)=\frac{4}{3^3}\left(\pi\frac{T_H}{m}\right)^4 e^{m/T_H}\,.
\en
The spectral density as a function of the mass $\tilde{\rho}(m)$ is defined via
\eq
\tilde{\rho}(m)\, \dd m=\rho(n)\,\dd n\,.
\label{rhotilde}
\en 
Using eq.~(\ref{MN}), one gets:
\eq
\dd n=\frac{3 m \dd m}{4\pi^2 (D-2) T_H^2}\,,
\label{dN}
\en
thus in $D=2+1$ dimensions one obtains:
\eq
\tilde\rho(m)=\frac{\pi^2}{9 T_H}\left( \frac{T_H}{m} \right)^3 e^{m/T_H}\,.
\en
The generalization to arbitrary $D=d+1$ is straightforward: for a closed string in $D$ spacetime dimensions, one finds:
\eq
\rho_{D}(n)=12(D-2)^{{D}}\left( \frac{\pi T_H}{3m} \right)^{D+1}
e^{m/T_H}\,.
\en
Combining this expression with eq.~(\ref{rhotilde}) and eq.~(\ref{dN}), one obtains:
\eq
\tilde{\rho}_D(m)= \frac{ (D-2)^{{D}-1} }{m} \left( \frac{\pi T_H}{3m} \right)^{D-1} e^{m/T_H}\,.
\label{rhoD}
\en

\end{document}